\documentclass[journal=jpcafh,manuscript=article,layout=twocolumn,email=false]{achemso}
\usepackage[utf8]{inputenc}
\usepackage{mathrsfs}
\usepackage{amssymb}
\usepackage{amsfonts}
\usepackage{amstext}
\usepackage{amsopn}
\usepackage{amsxtra}
\usepackage{mathrsfs}
\usepackage{dsfont}
\usepackage{graphicx}
\usepackage{pdfpages}

\newcommand{\R}{\mathbb{R}}

\newcommand{\N}{\mathbb{N}}

\newcommand{\ii}{\infty}
\newcommand\1{{\ensuremath {\mathds 1} }}

\newcommand{\rd}{{\rm d}}

\newcommand{\br}{\mathbf{r}}
\newcommand{\bR}{\mathbf{R}}
\newcommand\pscal[1]{{\ensuremath{\left\langle #1 \right\rangle}}}

\newcommand{\tr}{{\rm Tr}\,}
\renewcommand{\geq}{\geqslant}
\renewcommand{\leq}{\leqslant}

\usepackage{color}
\newcommand{\nn}{\nonumber}

\newtheorem{theorem}{Theorem}

\title[Non-convexity in $N$ of the ground state energy]{Ground state energy is not always convex in the number of electrons}

\author{Simone Di Marino}
\affiliation{Universit\`{a} di Genova, DIMA, MaLGa, Via Dodecaneso 35, 16146 Genova, Italy}

\author{Mathieu Lewin}
\email{mathieu.lewin@math.cnrs.fr}
\affiliation{CNRS \& CEREMADE, Universit\'e Paris-Dauphine, PSL University, 75016 Paris, France}

\author{Luca Nenna}
\affiliation{Universit\'e Paris-Saclay, CNRS, Laboratoire de math\'ematiques d'Orsay, ParMA, Inria Saclay, 91405, Orsay, France.}

\date{\today}

\SectionNumbersOn

\begin{document}
%%%%%%%%%%%%%%%%%%%%%%%%%%%%%%%%%%%%%%%%%%%%%%%%%%%%%%%%%%%%%%%%%%%%%%%%%%%%%%
%%%%%%%%%%%%%%%%%%%%%%%%%%%%%%%%%%%%%%%%%%%%%%%%%%%%%%%%%%%%%%%%%%%%%%%%%%%%%%

\begin{abstract}
We provide the first counterexample showing that the ground state energy of electrons in an external Coulomb potential is not always a convex function of the number of electrons. This convexity has been conjectured for decades and plays an important role in quantum chemistry. Our counterexample involves six nuclei with small fractional charges placed far apart. The ground state energy of 3 electrons is shown to be higher than the average of the energies for 2 and 4 electrons. We also show that the nuclei can bind 2 or 4 electrons, but not 3. This article raises the question of whether the energy convexity really holds for all possible molecules (with nuclei of integer charge).

\medskip

\footnotesize \noindent\copyright\;2024 the authors.
\end{abstract}

\maketitle

\medskip

\section{Introduction}

It seems to be an experimental fact that in an atom or molecule the core electrons are more tightly bound to the nuclei than the valence electrons. This property is measured by the ionization energy. Namely, if we remove the electrons from an atom one by one, the corresponding energy cost is seen to increase monotonically, which is a manifestation of the increasing difficulty of extracting the electrons. In this paper we will give the first example of a Coulomb system where the latter property fails. In our example, the third electron is more loosely bound to the nuclei than the fourth. This is certainly contrary to our intuition and, in fact, contradicts one of the most famous mathematical conjectures in quantum chemistry.

Let us first set up the stage. We consider $N$ electrons submitted to an external potential $V$. Their ground state energy is defined by
\begin{equation}
 E[V,N]:=\inf_{\Psi}\langle\Psi|H^V_N|\Psi\rangle
 \label{eq:EVN}
\end{equation}
with $\Psi$ the $N$-electron wavefunction and with the Hamiltonian operator
\begin{equation}
 H^V_N=\sum_{j=1}^N-\frac12\nabla^2_{\br_j}+\sum_{j=1}^NV(\br_j)+\!\!\sum_{1\leq j<k\leq N}\frac{1}{|\br_j-\br_k|},
 \label{eq:HVN}
\end{equation}
in atomic units. We allow for a rather general Coulomb external potential $V$ of the form
\begin{equation}
V(\br)=-\sum_m \frac{z_m}{|\br-\bR_m|},
 \label{eq:V_Coulomb}
\end{equation}
where the $z_m>0$ are not necessarily assumed to be integers, as they should be for real nuclei in atomic units. Since $V$ is given and only the number $N$ of electrons will vary, we do not take the nuclear repulsion into account in the energy.

A famous open problem is to rigorously prove that the map $N\mapsto E[V,N]$ is \textbf{convex}, which means
\begin{equation}
 E[V,N]-E[V,N-1]\leq E[V,N+1]-E[V,N],
 \label{eq:conjecture}
\end{equation}
for every $N\geq1$, with the convention that $E[V,0]=0$ when there are no electrons. The two terms in~\eqref{eq:conjecture} are non-positive since adding an electron can only decrease the ground state energy. In the worse case the additional electron escapes to infinity and the energy does not change. The absolute value $|E[V,N]-E[V,N-1]|$ is the \emph{ionization energy} mentioned above, that is, the energy cost to remove one electron from a system of $N$ electrons in their ground state. The inequality~\eqref{eq:conjecture} thus means that the ionization energy is non-increasing in $N$ and the interpretation is that the core electrons are more tightly bound than the valence electrons. The property~\eqref{eq:conjecture} can equivalently be rephrased as
\begin{equation}
E[V,N]\leq\frac{E[V,N-1]+ E[V,N+1]}2.
 \label{eq:conjecture2}
\end{equation}
In chemistry, the difference
\begin{equation*}
\frac{E[V,N-1]+ E[V,N+1]}2-E[V,N]
\end{equation*}
is often called the \emph{absolute hardness}\cite{ParPea-83} and the convexity property~\eqref{eq:conjecture2} just means the latter is non-negative.

The property~\eqref{eq:conjecture} is of fundamental importance in several situations of practical interest, including chemical reactivity theory\cite{ParPea-83,CoWas-07}, molecular partitioning\cite{EllBurCohWas-10}, degenerate ground states\cite{YanZhaAye-00}, molecular dissociation\cite{BurLinOre-23} and the formation of gaps in molecules and solids\cite{CohMorYan-08,ZheCohMorHuYan-11}. It has thus been assumed for a long time that~\eqref{eq:conjecture} should hold quite generally.

\begin{figure}[t]
\includegraphics[width=\linewidth]{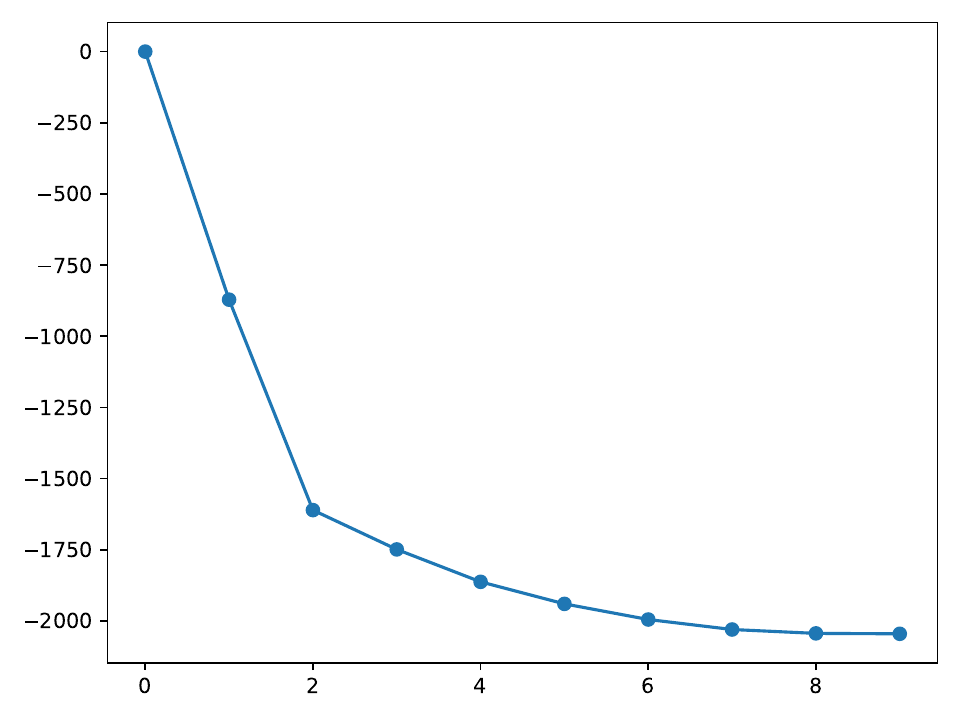}
\caption{Experimental ground state energy (eV) of Oxygen as a function of the number $N$ of electrons, with data from the NIST database\cite{NIST_database}. A similar graph can be found in Parr-Yang\cite{ParYan-94}.\label{fig:oxygen}}
\end{figure}

The importance of~\eqref{eq:conjecture} was mentioned for the first time in the context of Density Functional Theory (DFT) by Perdew, Parr, Levy and Balduz\cite{PerParLevBal-82} in 1982. In Chaper~4 of Ref.~\citenum{ParYan-94}, Parr and Yang stated the conjecture explicitly for atoms and molecules and provided experimental evidence for Carbon and Oxygen atoms (Figure~\ref{fig:oxygen}). More recent data largely confirm these findings\cite{NIST_database}. The conjecture~\eqref{eq:conjecture} is also explicitly stated in the mathematical physics literature. It is \emph{Question 7} in the famous 1983 article by Lieb\cite{Lieb-83b} on the foundations of DFT and it appeared a year later as \emph{Problem 10A} in a celebrated list of open questions by Simon.\cite{Simon-84} The conjecture is also mentioned on page~229 of Ref.~\citenum{LieSei-09}, in Conjecture~3 of Ref.~\citenum{Nam-22} and in Eq.~(3.18) of Ref.~\citenum{LewLieSei-23_DFT}.

There are very few situations where~\eqref{eq:conjecture} is rigorously known to hold. The first is for $N=1$, since after removing the Coulomb repulsion between two electrons we obtain
$E[V,2]\geq 2E[V,1]$.
The second case is for non-interacting systems\cite{PerParLevBal-82}. The only non-trivial example where~\eqref{eq:conjecture} could be proved rigorously is that of lithium-like atoms\cite{BriDucHog-87}.

It has been known for some time that if Conjecture~\eqref{eq:conjecture} is valid, this must be due to the particular nature of the Coulomb repulsion between the electrons. In fact, Lieb\cite{Lieb-83b} provided a counterexample for $N=2$, with classical hard spheres hopping on 4 sites in space. Lieb's construction was recently generalized by Ayers\cite{Ayers-24} to the Riesz interaction potential $|\br_j-\br_k|^{-s}$ with $s>2\log 2/\log3\approx 1.26$.
Hard spheres are recovered in the limit $s\to\ii$. Although $1.26$ is quite close to 1, no counterexample is known at the present for Coulomb. Other negative results concern particles interacting with a repulsive harmonic interaction in a harmonic trap $V$\cite{Levy-Leblond-68,PhiDav-83}, the 1D Hubbard model\cite{FyeMarSca-90} or a system in nuclear physics.\cite{BlaRip-85}

In this article we provide the \textbf{very first counterexample to Conjecture~\eqref{eq:conjecture} for the Coulomb interaction}. Similarly as for Lieb's counterexample\cite{Lieb-83b}, we first consider classical electrons hopping on finitely many sites. Then we extend the counterexample to the quantum case by a perturbation argument.

Our counterexample relies on a recent work\cite{MarLewNen-22_ppt}, where we studied the formalism of grand-canonical DFT\cite{PerParLevBal-82} for classical electrons. We found there a situation where the grand-canonical universal density functional does not coincide with its canonical counterpart. In the next section we explain that this is equivalent to the failure of the convexity property~\eqref{eq:conjecture}.

%%%%%%%%%%%%%%%%%%%%%%%%%%%%%%%%%%%%%%%%%%%%%%%%%%%%%%%%%%
%%%%%%%%%%%%%%%%%%%%%%%%%%%%%%%%%%%%%%%%%%%%%%%%%%%%%%%%%%
\section{Convexity and grand-canonical DFT}
Let us introduce the grand-canonical version of the ground state energy~\eqref{eq:EVN}.\cite{GyfHat-68} We fix the \emph{average} number of electrons $\lambda$, but do not assume that we have a canonical state. In particular, $\lambda$ need not be an integer. We obtain the grand-canonical energy
\begin{equation}
E_{\rm GC}[V,\lambda]=\inf_{\substack{(\Psi_n)_{n\geq0}\\\sum_{n\geq0}p_n=1\\ \sum_{n\geq1}np_n=\lambda}}\left\{\sum_{n\geq1}p_n\pscal{\Psi_n|H^V_n|\Psi_n}\right\}
\end{equation}
The coefficients $p_n$'s are the probabilities to have $n$ electrons. For each $n$, the lowest energy is reached by taking for $\Psi_n$ an $n$-electron ground state, leading to the simpler formula
\begin{equation}
E_{\rm GC}[V,\lambda]=\inf_{\substack{\sum_{n\geq0}p_n=1\\ \sum_{n\geq1}np_n=\lambda}}\left\{\sum_{n\geq1}p_n\,E[V,n]\right\}.
\end{equation}
The latter means that the grand-canonical energy is the \emph{convex hull} of the canonical energy $n\mapsto E[V,n]$ (extended into straight line between the integers).\cite{PerParLevBal-82,AyeLev-18,LewLieSei-23_DFT} In other words, it is the \emph{largest convex function below it}. Of course, the convex hull will do nothing when the function is already convex and therefore we obtain the following result.

\begin{theorem}\label{thm:GC}
Let us fix an arbitrary external potential~$V$ of the form~\eqref{eq:V_Coulomb}. The convexity property~\eqref{eq:conjecture} holds for this $V$ and all $N\in\N$, if and only if the grand-canonical and canonical ground state energies coincide for any number of electrons:
\begin{equation}
E_{\rm GC}[V,N]=E[V,N],\qquad \forall N\in\N.
\label{eq:E=EVGC}
\end{equation}
If this is the case, we obtain for a fractional average number of electrons $N+t$ with $0<t<1$
$$E_{\rm GC}[V,N+t]=(1-t)E[V,N]+tE[V,N+1].$$
\end{theorem}

Our next step will be to reformulate the equality~\eqref{eq:E=EVGC} by looking at the Legendre transform of the two functions $V\mapsto E[V,N]$ and $V\mapsto E_{\rm GC}[V,N]$. This is known as Lieb's convex formulation of Density Functional Theory\cite{Lieb-83b,HelTea-22,LewLieSei-23_DFT}. We recall that Lieb's universal functional\cite{Lieb-83b} is defined by
\begin{equation}
F[\rho]:=\inf_{\Gamma\mapsto\rho}\tr(H^0_N\Gamma).
 \label{eq:F_L}
\end{equation}
It provides the lowest kinetic plus interaction energy of a system of electrons in a mixed state~$\Gamma$, under the constraint that the density is exactly equal to the given $\rho(\br)$ everywhere. The method of constrained search for mixed states implies\cite{Levy-79,Lieb-83b}
\begin{equation}
E[V,N]=\inf_{\substack{\rho\\ \int_{\R^3}\rho=N}}\left\{F[\rho]+\int_{\R^3}\rho(\br)V(\br)\,\rd\br\right\}.
\label{eq:EN_FL}
\end{equation}
On the other hand, the Legendre-Fenchel duality theorem states that\cite{Lieb-83b}
\begin{equation}
F[\rho]=\sup_V\left\{E[V,N]-\int_{\R^3}\rho(\br)V(\br)\,\rd\br\right\},
\label{eq:FL_EN}
\end{equation}
with $N=\int_{\R^3}\rho(\br)\,\rd\br$. In other words, the two functionals $V\mapsto E[V,N]$ and $\rho\mapsto F[\rho]$ are Legendre-dual. This implies that all the information about one functional is contained in the other, and \emph{vice versa}.

We have the exact same situation in the grand-canonical case\cite{PerParLevBal-82,AyeLev-18,LewLieSei-23_DFT}. The grand-canonical universal functional is defined by
\begin{equation}
F_{\rm GC}[\rho]:=\inf_{\substack{(\Gamma)_{n\geq0}\\ \sum_{n\geq0}p_n=1\\ \sum_{n\geq1}p_n\rho_{\Gamma_n}=\rho}}\left\{\sum_{n\geq1}p_n\tr(H^0_n\Gamma_n)\right\}.
 \label{eq:F_GC}
\end{equation}
It satisfies the constrained search
\begin{equation}
E_{\rm GC}[V,\lambda]=\!\!\inf_{\substack{\rho\\ \int_{\R^3}\rho=\lambda}}\!\left\{F_{\rm GC}[\rho]+\int_{\R^3}\rho(\br)V(\br)\,\rd\br\right\}
\label{eq:EN_FL_GC}
\end{equation}
and the duality principle
\begin{equation}
F_{\rm GC}[\rho]=\sup_V\left\{E_{\rm GC}[V,\lambda]-\int_{\R^3}\rho(\br)V(\br)\,\rd\br\right\}.
\label{eq:FL_EN_GC}
\end{equation}
As a side remark we mention that $F$ and $F_{\rm GC}$ are well-defined only for representable densities, that is, those having a finite von Weizäcker energy $\int_{\R^3}|\nabla\sqrt\rho(\br)|^2\,\rd\br<\ii$\cite{Harriman-81,Lieb-83b}. If this is not the case, the convention is to take $F[\rho]=F_{\rm GC}[\rho]=+\ii$.

From the above duality formulas, it is clear that the equality $E[V,N]=E_{\rm GC}[V,N]$ for all $N$ and all $V$ is equivalent to the equality $F[\rho]=F_{\rm GC}[\rho]$ for all $\rho$. Using Theorem~\ref{thm:GC}, we obtain the following statement.

\begin{theorem}\label{thm:FL=FGC}
The convexity property~\eqref{eq:conjecture} holds for all $N\in\N$ and all $V$ of the form~\eqref{eq:V_Coulomb}, if and only if we have $F_{\rm GC}[\rho]= F[\rho]$ for all $\rho$ such that $\int_{\R^3}\rho(\br)\rd \br$ is an integer.
\end{theorem}

Since the validity of the convexity conjecture~\eqref{eq:conjecture} is equivalent to the equality of two universal functionals in DFT, we can focus on finding a density $\rho$ for which $F[\rho]> F_{\rm GC}[\rho]$. In the low density regime (or equivalently for strong interactions) the two functionals converge to their classical counterparts. It would thus be sufficient to find a counterexample for classical electrons. This is what was accomplished in our recent work\cite{MarLewNen-22_ppt}. But let us first quickly mention a positive result, showing that the space dimension plays a crucial role here.

%%%%%%%%%%%%%%%%%%%%%%%%%%%%%%%%%%%%%%%%%%%%%%%%%%%%%%%%%%
%%%%%%%%%%%%%%%%%%%%%%%%%%%%%%%%%%%%%%%%%%%%%%%%%%%%%%%%%%
\section{Convexity for classical electrons on a line}

In Section 5 of Ref.~\citenum{MarLewNen-22_ppt}  we studied the equivalent of the grand-canonical functional $F_{\rm GC}$ for \emph{classical} electrons evolving on a line. The latter can be taken to be the $x$ axis without any loss of generality. This is a situation that can be completely solved\cite{ColPasMar-15}. The proof is however involved and we will not attempt to describe it here. It relies both on the positivity and on the convexity of the Coulomb interaction. The final result is that for $\int_{\R}\rho(x)\,\rd x=N\in\N$ there must be exactly $N$ electrons in the system (hence the equality of the canonical and grand-canonical functionals) and that these $N$ electrons are ``strictly correlated'', in the sense that the position of all of them is entirely fixed by that of one of them only\cite{Seidl-99,SeiGorSav-07,SeiMarGerNenGieGor-17,SeiBenKooGor-22}.

By Theorem~\ref{thm:FL=FGC}, the coincidence of the canonical and grand-canonical classical density functionals implies immediately the following.

\begin{theorem}\label{thm:1D}
The convexity property~\eqref{eq:conjecture} holds for any potential $V$ and all $N\in\N$, for classical electrons constrained to stay on a line.
\end{theorem}

To our knowledge, this is the first positive result on the convexity property~\eqref{eq:conjecture}, for a non-trivial interacting system. We will see another example below.

%%%%%%%%%%%%%%%%%%%%%%%%%%%%%%%%%%%%%%%%%%%%%%%%%%%%%%%%%%
%%%%%%%%%%%%%%%%%%%%%%%%%%%%%%%%%%%%%%%%%%%%%%%%%%%%%%%%%%
\section{Classical electrons hopping on finitely many sites}

Let us now turn to the main result of the paper, that is, the construction of a counterexample to the convexity property~\eqref{eq:conjecture}.

We consider $N$ classical electrons that are only allowed to hop on $K$ distinct sites $\bR_1,...,\bR_K\in\R^3$. We will start by explaining how to formulate the problem for an arbitrary number of sites $K$. Then we will quickly discuss the case of $K=4$ points that was considered by Lieb\cite{Lieb-83b} and Ayers\cite{Ayers-24}, before we provide the sought-after counterexample for $K=6$.

\subsection{Formulation of the problem}
For a subset of indices $I\subset\{1,...,K\}$, we call
\begin{equation}
c_I=\sum_{\substack{1\leq j<k\leq K\\ j,k\in I}}\frac{1}{|\bR_j-\bR_k|}
\label{eq:def_c_I}
\end{equation}
the total Coulomb potential when there is one electron placed at each $\bR_i$ with $i\in I$. The convention is that $c_I=0$ if $I$ is the empty set or contains only one index. For instance $c_{\{1,2\}}=1/|\bR_1-\bR_2|$. There is no need to consider the case where two electrons are at the same place, which yields an infinite Coulomb energy.  A classical state is simply a probability distribution on the electronic configurations, that is, a collection $0\leq p_I\leq1$ with $\sum_I p_I=1$. The corresponding average Coulomb energy is
\begin{equation}
\sum_{I\subset\{1,...,K\}} p_Ic_I.
\label{eq:classical_energy}
\end{equation}
In the canonical case we assume that the $p_I$'s are non-zero only for the $I$'s of cardinality $\#I=N$. In the grand-canonical case all the $I$'s are allowed. The electronic density equals
\begin{equation}
\rho=\sum_{k=1}^K\rho_k\,\delta_{\bR_k},\qquad \rho_k=\sum_{k\in I}p_I.
 \label{eq:classical_density}
\end{equation}
We denote by $F^{\rm cl}[\rho]$ and $F^{\rm cl}_{\rm GC}[\rho]$ the canonical and grand-canonical classical Coulomb energies at fixed density $\rho$. Those are obtained by minimizing~\eqref{eq:classical_energy} at fixed density~\eqref{eq:classical_density}, either over canonical or grand-canonical probabilities $p_I$'s.

The dual external potential problem requires working with $V$'s of the particular form
\begin{equation}
V(\br)=\begin{cases}
v_k&\text{if $\br=\bR_k$}\\
+\ii&\text{if $\br\notin\{\bR_1,...,\bR_K\}$,}
\end{cases}
\label{eq:classical_V}
\end{equation}
in order to force the electrons to hop on the sites $\bR_k$. Let us call $E^{\rm cl}[V,N]$ and $E^{\rm cl}_{\rm GC}[V,N]$ the canonical and grand-canonical energies in an external potential $V$ of the form~\eqref{eq:classical_V}, for instance
\begin{equation}
E^{\rm cl}[V,N]=\min_{\substack{I\subset\{1,...,K\}\\ \#I=N}}\bigg\{c_I+\sum_{i\in I}v_i\bigg\}.
\label{eq:CVN}
\end{equation}
% The situation is similar in the grand-canonical case.

We have the same duality properties as we had in~\eqref{eq:EN_FL}--\eqref{eq:FL_EN} and~\eqref{eq:EN_FL_GC}--\eqref{eq:FL_EN_GC} in the quantum case, e.g.,
\begin{equation}
F^{\rm cl}_{\rm GC}[\rho]=\sup_{v_1,...,v_K}\left\{E^{\rm cl}_{\rm GC}[V,N]-\sum_{k=1}^Kv_k\rho_k\right\}.
 \label{eq:dual_GC_cl}
\end{equation}
From these formulas, we conclude that a statement similar to Theorem~\ref{thm:FL=FGC} holds for our classical electrons hopping on the sites $\bR_1,...,\bR_K$. Our goal will thus be to find a density $\rho=\sum_{k=1}^K\rho_k\,\delta_{\bR_k}$ such that $F^{\rm cl}_{\rm GC}[\rho]<F^{\rm cl}[\rho]$.

\subsection{Convexity for $K=4$ sites}
Lieb\cite{Lieb-83b} found a counterexample to the convexity property for hard-spheres hopping on $K=4$ sites. Ayers\cite{Ayers-24}
recently extended this result to the Riesz potential $|\br_j-\br_k|^{-s}$ for $s>2\log 2/\log3\approx 1.26$. A natural question is whether one can push $s$ down to 1 and get a counterexample for Coulomb. We can answer the question negatively. The energy is always convex when the electrons are constrained to hop on $K\leq 4$ sites!

\begin{theorem}[Convexity for $K\leq 4$]\label{thm:K4}
We have $F^{\rm cl}_{\rm GC}[\rho]=F^{\rm cl}[\rho]$ for all $\rho$ of the form $\rho=\sum_{k=1}^4\rho_k\,\delta_{\bR_k}$. This implies
\begin{equation}
 E^{\rm cl}[V,N]\leq\frac{E^{\rm cl}[V,N-1]+E^{\rm cl}[V,N+1]}2,
 \label{eq:convexity_K4}
\end{equation}
for all $N\in\N$ and all $V$ of the form~\eqref{eq:classical_V} with $K\leq 4$.
\end{theorem}

Note that $E^{\rm cl}[V,N]=+\ii$ for $N\geq K+1$, since then two electrons have to be at the same site. The inequality~\eqref{eq:convexity_K4} is thus obvious for $N\geq K$.

We quickly describe the proof of Theorem~\ref{thm:K4} in Appendix~\ref{app:proof_4points} below, using results from Ref.~\citenum{MarLewNen-22_ppt}. We emphasize that our argument seems to break down if $K=5$ or if the interaction is replaced by $|\br_j-\br_k|^{-s}$ for $s>1$.

\subsection{Counterexample for $K=6$ sites}

\begin{figure}
\includegraphics[width=\linewidth]{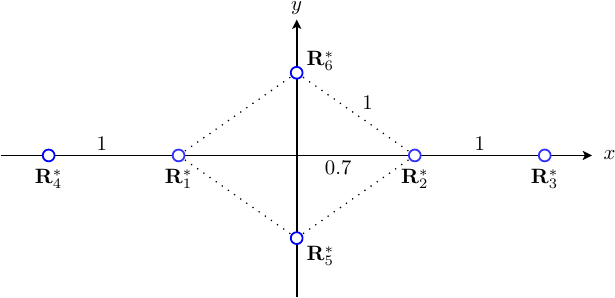}
\caption{Allowed locations of our classical electrons in the $(x,y)$ plane following Ref.~\citenum{MarLewNen-22_ppt}.\label{fig:diamond}}
\end{figure}

We finally turn to the counterexample. From Theorems~\ref{thm:1D} and~\ref{thm:K4}, we know that we have to take five points or more in two space dimensions. We will simply place $K=6$ points in a plane, that can be chosen to be the $(x,y)$--plane without loss of generality. We do not consider $K=5$ since we want to work at half-filling and thus need an even number of sites. A counterexample might well exist for $K=5$ too. Following Ref.~\citenum{MarLewNen-22_ppt}, we take
\begin{align}
\bR^*_1&=(-0.7,0,0),\quad &\bR^*_2&=(0.7,0,0),\nn\\
\bR^*_3&=(1.7,0,0),\quad &\bR^*_4&=(-1.7,0,0),\nn\\
\bR^*_5&=(0,\sqrt{0.51},0),\quad &\bR^*_6&=(0,-\sqrt{0.51},0),
\label{eq:Rj}
\end{align}
as displayed in Figure~\ref{fig:diamond}. In Ref.~\citenum{MarLewNen-22_ppt} we searched for counterexamples using a half-filling numerical optimization procedure with random positions for the 6 points. Several of the counterexamples we found looked a bit like the diamond of Figure~\ref{fig:diamond}. To simplify the computations, we then restricted the search to this particular shape and thus arrived at the mentioned values of $\bR^*_1,...,\bR^*_6$. We do not think there is anything special about this example, and many other counterexamples can be found. In fact, the points can be moved a little to suppress the symmetries, but not too much.\cite{MarLewNen-22_ppt}

Half-filling means that the density is constrained to be uniform, equal to 1/2 at each site,
\begin{equation}
\rho_{\text{h-f}}:= \frac12\sum_{k=1}^6\delta_{\bR_k^*},
\end{equation}
so that the average number of electrons is 3.
In this case the model is particle-hole symmetric, in addition to being invariant under reflections along the two axis. This simplifies the arguments quite a bit. After simple numerical computations, we found in Ref.~\citenum{MarLewNen-22_ppt} that the grand-canonical energy was strictly smaller than the canonical energy at half-filling:
\begin{multline*}
3.8778\approx F^{\rm cl}_{\rm GC}\left[\rho_{\text{h-f}}\right]<F^{\rm cl}\left[\rho_{\text{h-f}}\right]\approx 3.9157.
\end{multline*}
To be more precise, the grand-canonical optimizer corresponds to placing $2$ electrons at $\bR_1^*,\bR_2^*$ with probability $1/2$ and $4$ electrons at $\bR_3^*,...,\bR_6^*$ with probability $1/2$. The canonical problem is optimized by taking 3 electrons at $\bR_1^*,\bR_2^*,\bR_3^*$ or at $\bR_4^*,\bR_5^*,\bR_6^*$, with probability 1/2. Although the grand-canonical problem is lower, the energy gain is barely of  1\;\%.

Now, by the equivalent of Theorem~\ref{thm:FL=FGC} in the classical case, we conclude that there must exist one external potential $V$ of the form~\eqref{eq:classical_V}, for which the canonical energy is not convex in $N$. It remains to explain how to find the external potential $V$ in question.

The first potential we should try is the grand-canonical potential $V^{\rm GC}$ solving the dual problem~\eqref{eq:dual_GC_cl}. Indeed, duality theory tells us that one of the grand-canonical ground states for $V^{\rm GC}$ must have the exact density $\rho_\text{h-f}$ and the Coulomb energy $F^{\rm cl}_{\rm GC}[\rho_\text{h-f}]$. Since we know that the corresponding state is an average of a two-electron and a four-electron state, this is an indication that particle number is broken, as desired. We thus computed $V^{\rm GC}$ numerically, by solving Equation~\eqref{eq:dual_GC_cl} and found
\begin{equation}
\begin{cases}
v^{\rm GC}_1=v^{\rm GC}_2\approx-2.1731,\\
v^{\rm GC}_3=v^{\rm GC}_4\approx-1.3977,\\
v^{\rm GC}_5=v^{\rm GC}_6\approx -2.
\end{cases}
\label{eq:V_GC}
\end{equation}
The equalities are due to the space symmetry of the problem.
We expected a lack of convexity for the potential $V^{\rm GC}$ but the energies turn out to be all equal to each other:
\begin{align}
E^{\rm cl}[V^{\rm GC},2]&=E^{\rm cl}[V^{\rm GC},3]\nn\\
&=E^{\rm cl}[V^{\rm GC},4]\approx -3.6319.\label{eq:equality_GC}
\end{align}
In other words we obtain an exact equality in the bound~\eqref{eq:conjecture} we were trying to disprove!

\begin{figure}[t]
\includegraphics[width=9cm]{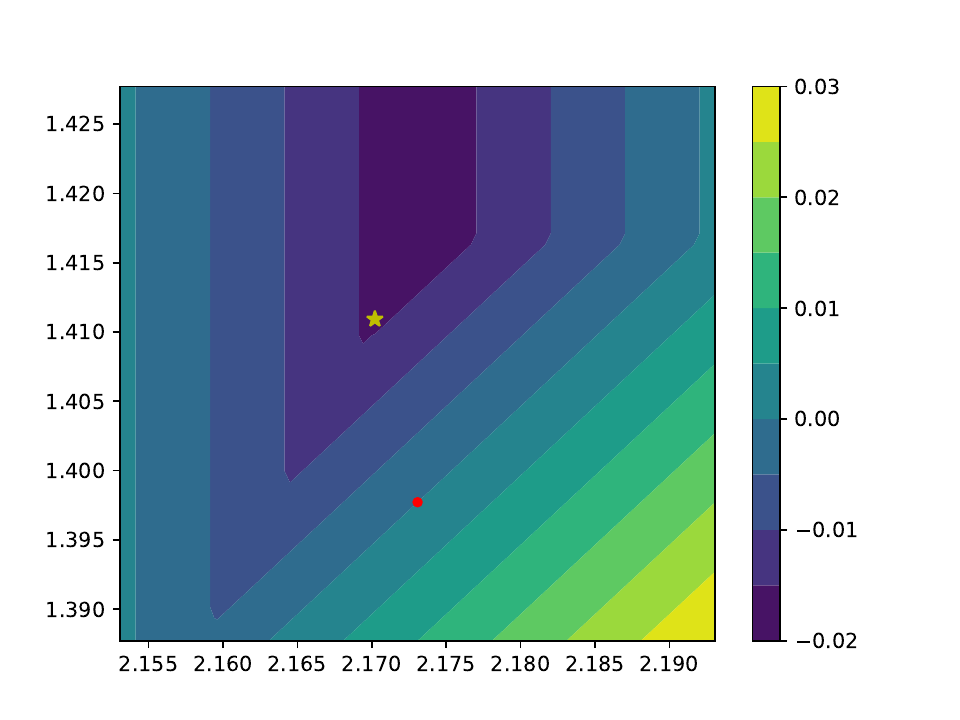}
 \caption{Level lines of the hardness $\eta[V]$ in~\eqref{eq:hardness} as a function of $|v_1|=|v_2|$ (horizontal axis) and $|v_3|=|v_4|$ (vertical axis), with $v_5=v_6=-2$ fixed. The red dot represents the grand-canonical dual potential $V^{\rm GC}$ in~\eqref{eq:V_GC} at which we have $\eta[V^{\rm GC}]=0$.  The yellow star represents the projection on the $(v_1,v_3)$ plane of the local minimum $V^*$ in~\eqref{eq:V_opt}, at which $\eta[V^*]<0$ and convexity is broken. The levels sets are exactly straight since the energy is linear in the $v_j$'s. \label{fig:level-lines}}
\end{figure}

We think that the equalities in~\eqref{eq:equality_GC} are an artefact of the particle-hole symmetry. To find a counterexample, we need to slightly go away from half-filling. We thus decided to numerically minimize the ``hardness''
\begin{equation}
V\mapsto \eta[V]:=\frac{E^{\rm cl}[V,2]+E^{\rm cl}[V,4]}2-E^{\rm cl}[V,3]
 \label{eq:hardness}
\end{equation}
in a neighborhood of the solution $V^{\rm GC}$ in~\eqref{eq:V_GC}. We found that $\eta$ was actually going down, as we had expected, in an appropriate direction. In Figure~\ref{fig:level-lines} we display its level lines in the $(v_1,v_3)$ plane, confirming that the grand-canonical potential $V^{\rm GC}$ of~\eqref{eq:V_GC} is exactly at the edge of a small valley where the hardness $\eta[V]$ becomes slightly negative.

Numerical minimization of the hardness $\eta[V]$ in~\eqref{eq:hardness} provided the new potential $V^*$
\begin{equation}
\begin{cases}
v^*_1=v^*_2=-2.1665,\\
v^*_3=v^*_4=-1.4109,\\
v^*_5=v^*_6= -1.9934.
\end{cases}
\label{eq:V_opt}
\end{equation}
Here we have truncated all the values at the fourth decimal. In the following we use these truncated values everywhere, and not the more precise approximation of the minimizer of~\eqref{eq:hardness} that we could compute numerically. It is enough to go to the fourth decimal to obtain a counterexample.

\begin{figure}[t]
 \includegraphics[width=7.5cm]{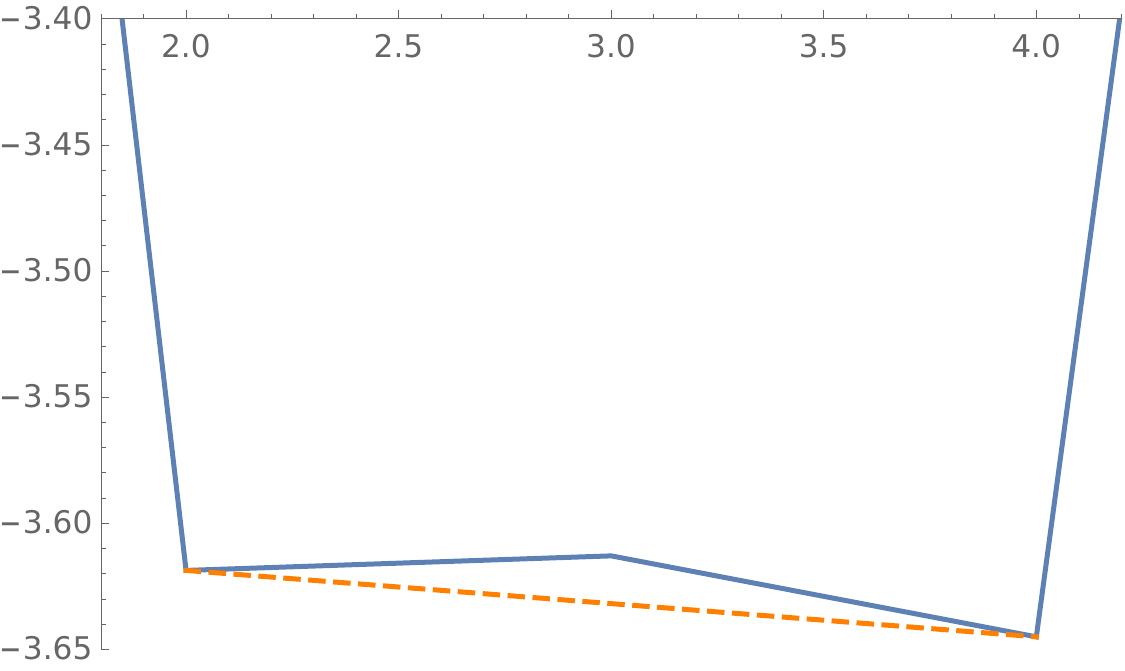}
 \caption{Classical energies $E^{\rm cl}[V^*,N]$ for $N\in\{2,...,4\}$, with $V^*$ as in~\eqref{eq:V_opt}. The function is not convex at $N=3$. \label{fig:energy_N}}
\end{figure}

Let us summarize the situation. We have $N$ electrons that are only allowed to be at the points $\bR_1,...,\bR_6$ in Figure~\ref{fig:diamond}, where they feel the potential given by~\eqref{eq:V_opt}. Looking at all the possibilities of placing the electrons as in~\eqref{eq:CVN}, we obtain the ground state energies $E^{\rm cl}[V^*,N]$ given in Table~\ref{tab:list_energies}. We also mention in the same table the optimal configurations for the electrons. Although we only display the first 4 digits, the energies can be computed up to machine precision. In particular, we find the claimed lack of convexity
\begin{multline}
-3.6319\approx  \frac{E^{\rm cl}[V^*,2]+E^{\rm cl}[V^*,4]}2\\
< E^{\rm cl}[V^*,3]\approx -3.6129.
\label{counter-energy}
\end{multline}
Note that the energy is even slightly going up at $N=3$, which is possible in such a confined system. The energies are plotted in Figure~\ref{fig:energy_N}.

To our knowledge, the potential $V^*$ defined in~\eqref{eq:classical_V} and~\eqref{eq:V_opt} provides the \textbf{very first example of a Coulomb system for which convexity-in-$N$ does not hold}.

\begin{table}[t]
\begin{tabular}{ccc}
$N$&$E^{\rm cl}[V^*,N]\approx$&minimizer\\
\hline
1 & -2.1665 & $\bR_1^*$\\
2 & -3.6187 & $\bR_1^*,\bR^*_2$\\
3 & -3.6129 & $\bR^*_4,\bR^*_5,\bR^*_6$\\
4 & -3.6450 & $\bR^*_3,...,\bR^*_6$\\
5 & -2.3949 & $\bR^*_2,...,\bR^*_6$\\
6 & -0.4304 & $\bR^*_1,...,\bR^*_6$\\
\end{tabular}

 \caption{Ground state energies and optimal configurations of the electrons for the external potential $V^*$ in~\eqref{eq:V_opt}.\label{tab:list_energies}}
\end{table}

We also tried to minimize the hardness~\eqref{eq:hardness} with respect to the positions $\bR_1^*,...,\bR_6^*$. We found a slightly asymmetric configuration of points where the hardness was a bit more negative. However, since the improvement was tiny, we did not pursue in this direction.

%%%%%%%%%%%%%%%%%%%%%%%%%%%%%%%%%%%%%%%%%%%%%%%%%%%%%%%%%%
%%%%%%%%%%%%%%%%%%%%%%%%%%%%%%%%%%%%%%%%%%%%%%%%%%%%%%%%%%
\section{Counterexample for quantum electrons}

To construct a counterexample with quantum electrons, we need first to force them to live close to the points $\bR_j^*$'s in~\eqref{eq:Rj} by creating very deep and narrow attractive wells. But then each well has a very negative energy, which we have to compensate by multiplying the Coulomb repulsion by a large constant. After rescaling everything, we see that this amounts to placing the nuclei very far apart.\cite{Ayers-24} We thus consider the following potential
\begin{equation}
V_\ell^*(\br)=-\sum_{j=1}^6\frac{z_j^*}{|\br-\ell \bR_j^*|},\quad z_j^*=\sqrt{\frac{2|v_j^*|}\ell}.
\label{eq:V_opt_nuclei}
\end{equation}
We recall once again that the $\bR_j^*$'s are given in~\eqref{eq:Rj} and the $v^*_j$'s in~\eqref{eq:V_opt}. The potential~\eqref{eq:V_opt_nuclei} corresponds to putting $6$ nuclei of very small fractional charges proportional to $1/\sqrt\ell$, far away from each other at a distance proportional to $\ell\gg1$. From a result of Lieb\cite{Lieb-84} we know that each of the 6 nuclei cannot bind more than 1 electron. The picture is, as we wanted, that we have to distribute our $N\leq 6$ electrons among the 6 nuclei, with at most one per nucleus. The ground state energy of each well equals $-(z_j^*)^2/2=v_j^*/\ell$ and it is comparable to the interaction energy between the electrons, so that we recover the classical picture to leading order in the limit $\ell\to\ii$.

Unlike the confined system studied in the previous section, the electrons can now choose to escape to infinity. A consequence is that the quantum energy $N\mapsto E[V_\ell^*,N]$ must be non-increasing. It cannot go up as it did in the previous classical example. The result is the following.

\begin{theorem}[Quantum counterexample]\label{thm:quantum}
The quantum energies in the potential $V^*_\ell$ of~\eqref{eq:V_opt_nuclei} satisfy, in the limit $\ell\to\ii$,
\begin{align*}
E[V^*_\ell,1]&=\frac{E^{\rm cl}[V^*,1]}{\ell}+o\left(\frac1\ell\right) \\
E[V^*_\ell,2]=E[V^*_\ell,3]&=\frac{E^{\rm cl}[V^*,2]}{\ell}+o\left(\frac1\ell\right) \\
E[V^*_\ell,N]=E[V^*_\ell,4]&=\frac{E^{\rm cl}[V^*,4]}{\ell}+o\left(\frac1\ell\right)
\end{align*}
for all $N\geq5$. The corresponding Hamiltonian $H^{V^*_\ell}_N$ in~\eqref{eq:HVN} admits a ground state for $N=1$, $N=2$ or $N=4$ electrons, but not for $N=3$ or $N\geq5$ electrons.
\end{theorem}

\begin{figure}[t]
 \includegraphics[width=7.5cm]{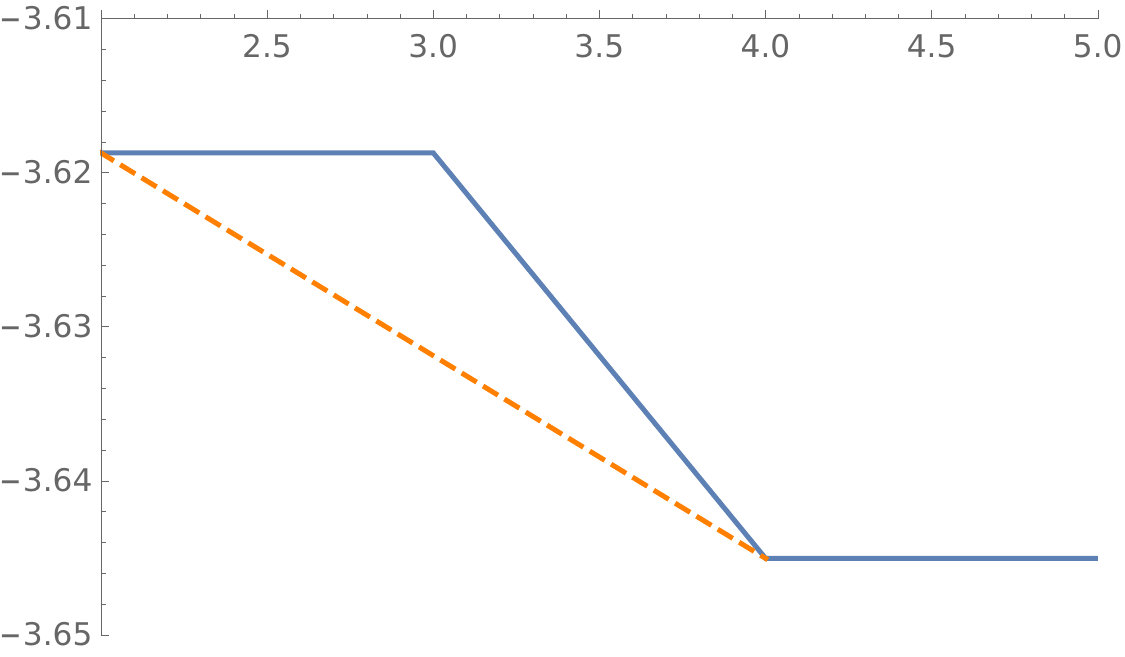}
 \caption{Quantum energies $\ell E[V^*_\ell,N]$ in the limit $\ell\to\ii$, for $N\in\{2,...,5\}$ and $V^*_\ell$ as in~\eqref{eq:V_opt_nuclei}. The function is not convex at $N=3$. \label{fig:energy_N_quantum}}
\end{figure}

We recall that the classical energies $E^{\rm cl}[V^*,N]$ are given in Table~\ref{tab:list_energies}. Although $E[V^*_\ell,3]$ is lower than in the classical case since the third electron prefers to escape to infinity, we still get the sought-after counterexample
$$\frac{E[V^*_\ell,2]+E[V^*_\ell,4]}2<E[V^*_\ell,3]=E[V^*_\ell,2]$$
for $\ell$ large enough, see Figure~\ref{fig:energy_N_quantum}. We also have the rather strange situation of a nuclear system that can bind 2 and 4 electrons but not 3.

Although Theorem~\ref{thm:quantum} is very intuitive in view of the results from the previous section, its rigorous proof is not so easy. The argument uses geometric localization techniques\cite{MorSim-80,Ruskai-82a,Sigal-82,LieSigSimThi-88,Lewin-11} and it is outlined in Appendix~\ref{app:proof_quantum}, for the convenience of the reader. Note that the result does not depend on the electrons between spin-$1/2$ particles. Since they end up being very far away from each other, the Pauli principle is not so important to leading order and the exact same theorem holds for bosons. The same result is also valid in Hartree-Fock theory.

We conclude this section with two remarks. First, one can add to the six nuclei any number of even smaller charges placed very far away to obtain a counterexample for $N=3$ and any $K\geq6$. By this procedure, one can also ensure that the total nuclear charge is an integer. Second, in our counterexample the grand-canonical problem for $N=3$ electrons ends up being a strict average of the $2$ and $4$ electron cases. It is also possible to construct a counterexample such that the grand-canonical $N$-electron problem is an average of the $N-J$ and $N+J$ canonical energies with $J$ very large. Indeed, in Theorem 3.16 of Ref.~\citenum{MarLewNen-22_ppt} we gave the example of $K=6^k$ points (a kind of multiscale copy of the 6 points in~\eqref{eq:Rj} at $k$ different scales), for which the grand-canonical classical problem for $N=6^k/2$ is an average of the $(6^k-2^k)/2$ and $(6^k+2^k)/2$ canonical problems. Hence the grand-canonical ground state has a number of electrons that varies a lot around the mean $N$, by $N^{\frac{\log2}{\log6}}\gg1$. This gives rise to a quantum counterexample by the same proof as for Theorem~\ref{thm:quantum}.

\section{Conclusion}

Based on results from our previous work on classical grand-canonical DFT\cite{MarLewNen-22_ppt} and a duality principle, we have provided the very first example of a system of classical electrons for which the ground state energy is not convex in the number of electrons. A perturbation argument then allowed us to construct a quantum counterexample, with 6 nuclei of very small fractional charges placed far apart. It is a very interesting open problem to find a system of real nuclei (that is, of integer charges) for which the convexity property~\eqref{eq:conjecture} fails.
We hope that this paper will stimulate further work in this direction.

\begin{acknowledgement}
This paper is dedicated to Trygve Helgaker on the occasion of his 70th birthday.
L.N. benefited from the support of the FMJH Program PGMO,  from H-Code, Universit\'e Paris-Saclay and from the ANR project GOTA (ANR-23-CE46-0001). S.D.M. benefited from the support of  MUR (PRIN project 202244A7YL and DipE 2023-2027 awarded to the DIMA (Unige), CUP D33C23001110001), Air Force (AFOSR project FA8655-22-1-7034) and Istituto Nazionale di Alta Matematica (he is a member of GNAMPA).

\end{acknowledgement}

\appendix

%%%%%%%%%%%%%%%%%%%%%%%%%%%%%%%%%%%%%%%%%%%%%
\section{Proof of Theorem~\ref{thm:K4}}\label{app:proof_4points}
We describe here the proof of Theorem~\ref{thm:K4}, using results from Ref.~\citenum{MarLewNen-22_ppt}. Recall that we wish to prove the equality $F^{\rm cl}_{\rm GC}[\rho]=F^{\rm cl}[\rho]$ for all densities $\rho=\sum_{k=1}^K\rho_k\delta_{\bR_k}$ with $K\leq4$ and $N:=\sum_{k=1}^K\rho_k\in\N$.

First, if $\rho_k>1$ for one $k$ (as must be the case if there are more electrons than sites, $N>K$) we obtain immediately $F^{\rm cl}_{\rm GC}[\rho]=F^{\rm cl}[\rho]=+\ii$. Another easy situation is when $\rho_k\in \{0,1\}$ for all $k$, which means that we have to put one electron at each of the sites where $\rho_k=1$, and thus get a canonical state. This covers the situation that $N=K$, for instance. Finally, if $N=1$, we have $F^{\rm cl}_{\rm GC}[\rho]=F^{\rm cl}[\rho]=0$. The corresponding ground state amounts to putting one electron at the $\bR_k$'s with probability $\rho_k$. We thus only have to look at $N=2$ and $N=3$.

The most complicated case is $N=2$, but it was handled in our recent work\cite{MarLewNen-22_ppt}. It is indeed proved in Thm.~3.15 of Ref.~\citenum{MarLewNen-22_ppt} that $F^{\rm cl}_{\rm GC}[\rho]=F^{\rm cl}[\rho]$ for all densities $\rho$ such that $\int_{\R^3}\rho(\br)\,\rd\br=2$. It is not necessary that $\rho$ is a finite combination of Dirac deltas. Note that this result really uses the Coulomb interaction. We do not have such a strong information if for instance the interaction is replaced by $|\br_j-\br_k|^{-s}$ for some $s>1$ (see Thm~3.9 in Ref.~\citenum{MarLewNen-22_ppt}).

It remains to treat the case $N=3$ and $K=4$. We can assume $\rho_k>0$ for all $k$ (if one $\rho_k$ vanishes we must put the three electrons at the three other sites). The same Thm.~3.15 of Ref.~\citenum{MarLewNen-22_ppt} specifies that any grand-canonical ground state must involve only $n=2,3,4$ electrons. In other words, we have $p_I=0$ if $I$ is empty or contains one index. Let us now argue by contradiction and assume that, for instance, $p_{\{1,2\}}>0$. Then we must have $p_{\{1,2,3,4\}}>0$ as well and, of course, $p_I<1$ for every configuration $I$. Next we modify the ground state by replacing
\begin{align*}
p_{\{1,2\}}&\rightsquigarrow p_{\{1,2\}}-t\\
p_{\{1,2,3,4\}}&\rightsquigarrow p_{\{1,2,3,4\}}-t\\
p_{\{1,2,3\}}&\rightsquigarrow p_{\{1,2,3\}}+t\\
p_{\{1,2,4\}}&\rightsquigarrow p_{\{1,2,4\}}+t
\end{align*}
for $t>0$ small enough. The indices were chosen so that the density is not affected by these modifications. Hence the energy must go up and we find
\begin{multline*}
0\leq t\big(c_{\{1,2,3\}}+c_{\{1,2,4\}}-c_{\{1,2\}}-c_{\{1,2,3,4\}}\big)\\
=-\frac{t}{|\bR_3-\bR_4|}.
\end{multline*}
This is definitely a contradiction and hence we have shown $p_{\{1,2\}}=0$. There was of course nothing special with $\{1,2\}$ and by permuting the labels we conclude that $p_{\{i,j\}}=0$ for every $i< j$. Since we have 3 electrons in average this implies $p_{\{1,2,3,4\}}=0$ and therefore our state must be canonical, as was claimed.

%%%%%%%%%%%%%%%%%%%%%%%%%%%%%%%%%%%%%%%%%%%%%
\section{Proof of Theorem~\ref{thm:quantum}}\label{app:proof_quantum}

\subsection{Asymptotic expansion}\label{sec:asymptotics}
In this subsection we discuss the proof of the asymptotic expansion
\begin{equation}
E[V^*_\ell,N]= \frac{\min_{n=1,...,N}E^{\rm cl}[V^*,n]}\ell+O(\ell^{-\frac54})
\label{eq:asymptotics}
\end{equation}
for all $N\geq1$. The error is a little bit more precise than that stated in the theorem, but it is probably not optimal.

First we show an upper bound, that is, we prove~\eqref{eq:asymptotics} with a $\leq$. Let $N\leq6$ and consider the optimal $\bR_{j_1},...,\bR_{j_N}$ for $E^{\rm cl}[V^*,N]$ given in Table~\ref{tab:list_energies}. We place one electron at each of these nuclei, in the hydrogenic ground state $\phi_{k,\ell}(\br,\sigma)=\frac1{\sqrt\pi}e^{-z^*_{j_k}|\br-\ell\bR^*_{j_k}|}\delta_{\sigma\uparrow}$. More precisely, we first orthonormalize the $\phi_{k,\ell}$'s by Gram-Schmidt with an exponentially small error in $\ell$ and we can then take as trial state the associated (spin-polarized) Slater determinant. We find
$$E[V^*_\ell,N]\leq \frac{E^{\rm cl}[V^*,N]}\ell+O(\ell^{-\frac32})$$
where the error in $\ell^{-3/2}$ comes from the interaction between each electron and the other nuclei. For $N=3$ we can first send one electron to infinity and we obtain the better upper bound
$$E[V^*_\ell,3]\leq E[V^*_\ell,2]\leq \frac{E^{\rm cl}[V^*,2]}\ell+O(\ell^{-\frac32}).$$
We argue similarly for $N\in\{5,6\}$. Finally, a theorem of Lieb\cite{Lieb-84} states that a system of $M$ nuclei of total (possibly fractional) charge $Z$ can never bind more than $2Z+M$ electrons. In our case the total nuclear charge goes to zero in the limit $\ell\to\ii$, hence our system can never bind more than $6$ electrons. We thus have $E[V^*_\ell,N]=E[V^*_\ell,6]$ all $N\geq7$, without any ground state. We have thus proved the upper bound in~\eqref{eq:asymptotics}, with the (probably optimal) error $O(\ell^{-3/2})$.

Let us now turn to the proof of the lower bound, using standard localization techniques.\cite{MorSim-80,Lewin-11} We pick a smooth radial function $\chi$ which equals 1 in the ball of radius 1 and vanishes outside of the ball of radius 2. We then define the localization function about the $j$-th nucleus by
$$\chi_{j,\ell}(\br)=\chi\left(\frac{\br-\ell \bR^*_j}{\ell^{3/4}}\right).$$
Since we know that the hydrogenic ground states live at the scale $\sqrt\ell$ we need to localize at a larger scale. On the other hand, we do not want to reach the distance $\sim \ell$ between the nuclei. Hence $\ell^{3/4}$ sounds like a good compromise. We define the complementary localization $\chi_{7,\ell}$ so as to obtain a partition of unity, $\sum_{j=1}^7\chi_{j,\ell}^2=1$. Plugging this partition in each variable $\br_j$ and expanding, we obtain a partition of unity in the $N$-electron space. This leads to
\begin{multline}
\big\langle\Psi\big|H^{V^*_\ell}_N\big|\Psi\big\rangle\\
=\sum_{j_1,...j_N=1}^7\Big\langle \chi_{j_1,...,j_N,\ell}\Psi\Big|H^{V^*_\ell}_N\Big|\chi_{j_1,...,j_N,\ell}\Psi\Big\rangle\\+O(\ell^{-\frac32})\label{eq:IMS}
\end{multline}
for any $N$-particle wavefunction $\Psi$, where
$$\chi_{j_1,...j_N,\ell}(\br_1,...,\br_N):=\chi_{j_1,\ell}(\br_1)\cdots\chi_{j_N,\ell}(\br_N)$$
localizes in the region where the $k$-th electron is on the support of $\chi_{j_k,\ell}$. The $O(\ell^{-3/2})$ comes from the gradient of the localization, due to the IMS formula\cite{CycFroKirSim-87}.

We then look at all the possibilities for the indices $j_1,...,j_7$, corresponding to all the ways of placing the $N$ electrons in the 7 regions. In each case we bound the energy from below, applying the following rules:

\smallskip

\noindent$\bullet$ The interaction between an electron close to a nucleus and another nucleus is of order $\ell^{-3/2}$;

\smallskip

\noindent$\bullet$ The interaction between an electron in the 7-th region and all the nuclei is at most of order $\ell^{-5/4}$;

\smallskip

\noindent$\bullet$ For a lower bound we can discard the kinetic energy of an electron in the 7-th region, as well as its repulsion with the other electrons;

\smallskip

\noindent$\bullet$ If we have two electrons or more close to one nucleus, the corresponding repulsion is bounded-below by $\frac1{4\ell^{3/4}}$, a positive term that dominates all the other terms and is also much larger than the desired $\min_{n=1,...,N}E^{\rm cl}[V^*,n]/\ell$;

\smallskip

\noindent$\bullet$ The repulsion between two electrons located close to $\ell\bR_j$ and $\ell\bR_k$ with $k\neq j$ can be bounded from below by
$$\frac1{\ell|\bR^*_j-\bR^*_k|+4\ell^{3/4}}=\frac1{\ell|\bR^*_j-\bR^*_k|}+O(\ell^{-\frac54});$$

\smallskip

\noindent$\bullet$ For the interaction between an electron and its nucleus we use the hydrogenic ground state energy in the operator form
$$-\frac12\nabla_\br^2-\frac{z_j^*}r\geq -\frac{(z_j^*)^2}2=\frac{v^*_j}\ell.$$

\smallskip

\noindent Applying all the above rules we obtain the operator bound
\begin{multline*}
\chi_{j_1,...,j_N,\ell}H^{V^*_\ell}_N\chi_{j_1,...,j_N,\ell}\\
\geq \left(\frac{\min_{n=1,...,N}E^{\rm cl}[V^*,n]}\ell+O(\ell^{-\frac54})\right)\chi_{j_1,...,j_N}^2.
\end{multline*}
Summing over $j_1,...,j_N$ leads to the desired lower bound, hence to the stated asymptotic expansion~\eqref{eq:asymptotics}.

\subsection{No binding for $N=3,5,6$}
Although we know that $E[V_\ell^*,3]$ and $E[V_\ell^*,2]$ have the same asymptotic expansion to leading order in the limit $\ell\to\ii$, this does not yet prove that $E[V_\ell^*,3]=E[V_\ell^*,2]$. We need to be more precise to obtain the exact equality. The difficulty is to prove that one of the three electrons feels a repulsive potential everywhere in space, that pushes it to infinity. The proof relies on a different localization technique\cite{Ruskai-82a,Sigal-82,CycFroKirSim-87,LieSigSimThi-88}. Its goal is to find which electron is the furthest away and at what distance. Once we know this distance we can use the electronic repulsion to control the localization errors. We only outline this proof for $N=3$, the argument is very similar for $N=5,6$.

Let $\eta:[0,\ii)\to[0,1]$ be a smooth non-decreasing function such that $\eta\equiv0$ on $[0,1]$ and $\eta\equiv1$ on $[2,\ii)$. Denote by $D(\br_1,\br_2,\br_3):=\max\{|\br_1|,|\br_2|,|\br_3|\}$ the distance to the origin of the furthest away electron. For a large enough constant $K$ (to be determined below), we introduce for $j=1,2,3$
$$F_j(\br_1,\br_2,\br_3):=\eta\left(\frac{D}{K\ell}\right)\eta\left(\frac{2|\br_j|}{D}\right).$$
The first factor imposes that $D$ is large enough whereas the second requires that the $j$-th electron is located at a distance comparable to $D$. We have $\sum_{j=1}^3F_j^2\geq \eta(\frac{D}{K\ell})^2$
since in the sum there exists one $j$ for which $|\br_j|=D$ and $\eta(2)=1$. We then let $F_0:=1-\eta(\frac{D}{K\ell})$
and obtain $\frac12\leq \sum_{j=0}^3F_j^2\leq3$. This allows us to introduce the new partition of unity
$$\Xi_j:=\frac{F_j}{\left(\sum_{k=0}^3F_k^2\right)^{1/2}}.$$
Outside of the set of zero measure where two electrons are the furthest away, it can be seen that the function $F_j$ is differentiable, with
$$|\nabla F_j|\leq \sqrt{C}\frac{\1(D\geq K\ell)}{D}$$
for some explicit constant $C$ that is not important for our argument. We have a similar bound on $|\nabla\Xi_j|$.
The IMS formula provides
\begin{multline}
\pscal{\Psi,H^{V^*_\ell}_3\Psi}\\
\geq \sum_{j=0}^3\pscal{\Xi_j\Psi\bigg|H^{V^*_\ell}_3-\frac{C\1(D\geq K\ell)}{D^2}\bigg|\Xi_j\Psi}.
\label{eq:localized_furthest}
\end{multline}
The localization function $\Xi_1$ corresponds to the first electron being far away at a distance at least $D/2\geq K\ell/2$. By definition of $D$ we have $|\br_1-\br_m|\leq 2D$ for $m=2,3$, hence we can bound the repulsion between the first electron and the other two by
$$\frac1{|\br_1-\br_2|}+\frac1{|\br_1-\br_3|}\geq \frac1D.$$
On the other hand, we can estimate the nuclear attraction using
$$|\br_1-\ell\bR^*_m|\geq |\br_1|-2\ell\geq D/4+(K/4-2)\ell,$$
since $|\bR^*_m|\leq 2$ by definition. We thus pick $K=8$ to obtain $|\br_1-\ell\bR^*_m|\geq D/4$. Using $\sum_{m=1}^6(2|v_m^*|)^{1/2}\leq 12$ we arrive at
\begin{align*}
&\Xi_1\left(H^{V^*_\ell}_3-\frac{C}{D^2}\right)\Xi_1\\
&\qquad \geq \Xi_1\left((H^{V^*_\ell}_2)_{\br_2,\br_3}+\frac1D-\frac{48}{D\sqrt\ell}-\frac{C}{D^2}\right)\Xi_1\\
&\qquad \geq \left(E[V^*_\ell,2]+\frac1{2D}\right)\Xi_1^2
\end{align*}
for $\ell$ large enough, since $D\geq K\ell=8\ell$. Here $(H^{V^*_\ell}_2)_{\br_2,\br_3}$ denotes the 2-electron Hamiltonian acting on the second and third electrons. The bound is exactly the same for $j=2,3$, by symmetry.

Next we look at the support of $\Xi_0$ where $D\leq 2K\ell=16\ell$, that is, all the electrons are at distance $\ell$ to the nucleus. On that set we just localize further, using the same $\chi_{j,\ell}$ as we did in Section~\ref{sec:asymptotics}. What we have gained here is that the repulsion between an electron in the 7-th region and an electron close to a nucleus can now be bounded from below by $1/(32\ell)$, a term that can be used to control the localization errors and the interaction with the other nuclei. In the case that we have no electron in the 7-th region, we instead use $E^{\rm cl}[V^*,3]>E^{\rm cl}[V^*,2]$ to also gain a positive term of order $1/\ell$. We end up with an estimate of the form
\begin{align*}
&\Xi_0\left(H^{V^*_\ell}_3-\frac{C\1(D\geq K\ell)}{D^2}\right)\Xi_0\\
&\qquad \geq \left(\frac{E^{\rm cl}[V^*,2]}{\ell}+\frac\alpha{\ell}+O(\ell^{-\frac65})\right)\Xi_0^2\\
&\qquad \geq \left(E[V^*_\ell,2]+\frac\alpha{2\ell}\right)\Xi_0^2
\end{align*}
for $\alpha=\min(1/32, E^{\rm cl}[V^*,3]-E^{\rm cl}[V^*,2])\approx 0.0058$ and all $\ell\gg1$. In the last line we used the asymptotic expansion~\eqref{eq:asymptotics}.

The two bounds prove that for any wavefunction $\Psi$
$$\pscal{\Psi\Big|H^{V^*_\ell}_3\Big|\Psi}\geq E[V^*_\ell,2]+\int_{\R^9}\frac{|\Psi|^2}{2\max(D,\ell\alpha^{-1})}.$$
Hence $E[V^*_\ell,3]\geq E[V^*_\ell,2]$, which implies equality due to the monotonicity. Also, there cannot exist a ground state $\Psi$ since the last term would have to vanish. This concludes the proof of Theorem~\ref{thm:quantum}.

%%%%%%%%%%%%%%%%%%%%%%%%%%%%%%%%%%%%%%%%%%
% \bibliographystyle{achemso}
% \bibliography{biblio}

\providecommand{\latin}[1]{#1}
\makeatletter
\providecommand{\doi}
  {\begingroup\let\do\@makeother\dospecials
  \catcode`\{=1 \catcode`\}=2 \doi@aux}
\providecommand{\doi@aux}[1]{\endgroup\texttt{#1}}
\makeatother
\providecommand*\mcitethebibliography{\thebibliography}
\csname @ifundefined\endcsname{endmcitethebibliography}
  {\let\endmcitethebibliography\endthebibliography}{}

\end{document}